\documentclass{bmcart}

\usepackage[utf8]{inputenc} 
\usepackage{graphicx}
\usepackage{latexsym}
\usepackage{amsfonts,amsmath,amssymb}
\usepackage{bm}
\usepackage{bbm}
\usepackage{epsfig}
\usepackage{relsize}
\usepackage{color}
\usepackage{hyperref}

\startlocaldefs

\definecolor{dred}{rgb}{.8,0.2,.2}
\definecolor{dyellow}{rgb}{.7,.7,.0}
\definecolor{ddred}{rgb}{.4,.0,.0}
\definecolor{dblue}{rgb}{.2,.2,.8}

\newcommand{\ket}[1]{\left | \, #1 \right \rangle}

\newcommand{\bra}[1]{\left \langle #1 \, \right |}

\newcommand{\an}[1]{\hat{#1}}

\newcommand{\dd}{\mathrm{d}}

\newcommand{\rrr}{\mathbf{r}}
\newcommand{\rr}{\mathrm{r}}

\newcommand{\fir}[1]{Fig.~\ref{#1}}
\newcommand{\secr}[1]{Sec.~\ref{#1}}

\endlocaldefs

\begin{document}

\begin{frontmatter}

\begin{fmbox}
\dochead{Research}


\title{What is a quantum simulator?}


\author[
   addressref={aff1,aff2,aff3,aff4},                   
   corref={aff1},                       
   email={tomi.johnson@physics.ox.ac.uk}   
]{\inits{THJ}\fnm{Tomi H} \snm{Johnson}}
\author[
   addressref={aff2,aff1,aff3}
]{\inits{SR}\fnm{Stephen R} \snm{Clark}}
\author[
   addressref={aff2,aff1,aff3}
]{\inits{D}\fnm{Dieter} \snm{Jaksch}}


\address[id=aff1]{
  \orgname{Centre for Quantum Technologies, National University of Singapore}, 
  \street{3 Science Drive 2},                     %
  \postcode{117543}                                
  \city{Singapore},                              
  \cny{Singapore}                                    
}
\address[id=aff2]{%
  \orgname{Clarendon Laboratory, University of Oxford},
  \street{Parks Road},
  \postcode{OX1 3PU}
  \city{Oxford},
  \cny{UK}
}
\address[id=aff3]{%
  \orgname{Keble College, University of Oxford},
  \street{Parks Road},
  \postcode{OX1 3PG}
  \city{Oxford},
  \cny{UK}
}
\address[id=aff4]{%
  \orgname{Institute for Scientific Interchange},
  \street{Via Alassio 11/c},
  \postcode{10126}
  \city{Torino},
  \cny{Italy}
}


\begin{artnotes}
\end{artnotes}

\end{fmbox}


\begin{abstractbox}

\begin{abstract} 
Quantum simulators are devices that actively use quantum effects to answer questions about model systems and, through them, real systems.
Here we expand on this definition by answering several fundamental questions about the nature and use of quantum simulators. Our answers address two important areas. First, the difference between an operation termed simulation and another termed computation. This distinction is related to the purpose of an operation, as well as our confidence in and expectation of its accuracy. Second, the threshold between quantum and classical simulations. Throughout, we provide a perspective on the achievements and directions of the field of quantum simulation.
\end{abstract}


\begin{keyword}
\kwd{Quantum}
\kwd{Simulation}
\kwd{Computation}
\kwd{Definition}
\kwd{Requirements}
\kwd{Perspective}
\end{keyword}

\begin{keyword}[class=AMS]
\kwd{03.65.-w}
\kwd{03.67.Ac}
\kwd{03.67.Lx}
\end{keyword}

\end{abstractbox}
%

\end{frontmatter}



\section{Introduction}
Simulating models of the physical world is instrumental in advancing scientific knowledge and developing technologies. Accordingly, the task has long been at the heart of science. For example, orreries have been used for millennia to simulate models of the motions of celestial objects~\cite{Brewster1830}. More recently, differential analysers or mechanical integrators were developed to solve differential equations modelling e.g.\ heat flow and transmission lines~\cite{Thomson1876,Cairns1944}. 

Unfortunately, simulation is not always easy. There are numerous important questions to which simulations would provide answers but which remain beyond current technological capabilities. These span a multitude of scientific research areas, from high-energy~\cite{Barcelo2011,Jordan2012}, nuclear, atomic~\cite{You2011} and condensed matter physics~\cite{Lewenstein2007,Lewenstein2012} to thermal rate constants~\cite{Lidar2009} and molecular energies~\cite{Wang2008,Whitfield2011} in chemistry~\cite{Kassal2011,Lu2012}.

An exciting possibility is that the first simulation devices capable of answering some of these questions may be quantum, not classical, with this distinction to be clarified below. The types of quantum hardware proposed to perform such simulations are as hugely varying as the problems they aim to solve: trapped ions \cite{Porras2004,Kim2010,Gerritsma2010,Lanyon2011,Blatt2012}, cold atoms in optical lattices \cite{Jaksch2005,Dalibard2011,Bloch2012}, liquid and solid-state NMR \cite{Peng2009,Peng2010,Du2010,Li2011,Zhang2011}, photons \cite{Angelakis2007,Angelakis2008,Lu2009,Lanyon2009,Peruzzo2010,Aspuru-Guzik2012}, quantum dots \cite{Manousakis2002,Smirnov2007,Byrnes2008}, superconducting circuits \cite{You2005,Clarke2008,You2011,Houck2012}, and NV centres~\cite{Wang2014}. At the time of writing, astonishing levels of control in proof-of-principle experiments (cf.\ the above references and citations within) suggest that quantum simulation is transitioning from a theoretical dream into a credible possibility.

Here we complement recent reviews of quantum simulation~\cite{Kendon2010,Cirac2012,Hauke2012,Schaetz2013,Buluta2009,Georgescu2013} by providing our answers to several fundamental but non-trivial and often contentious questions about quantum simulators, highlighting whenever there is a difference of opinion within the community. In particular, we discuss how quantum simulations are defined, the role they play in science, and the importance that should be given to verifying their accuracy.

\section{What are simulators?}
\label{sec:whatsim}
\begin{figure}[tbp]
\begin{center}
  \includegraphics[width=85mm]{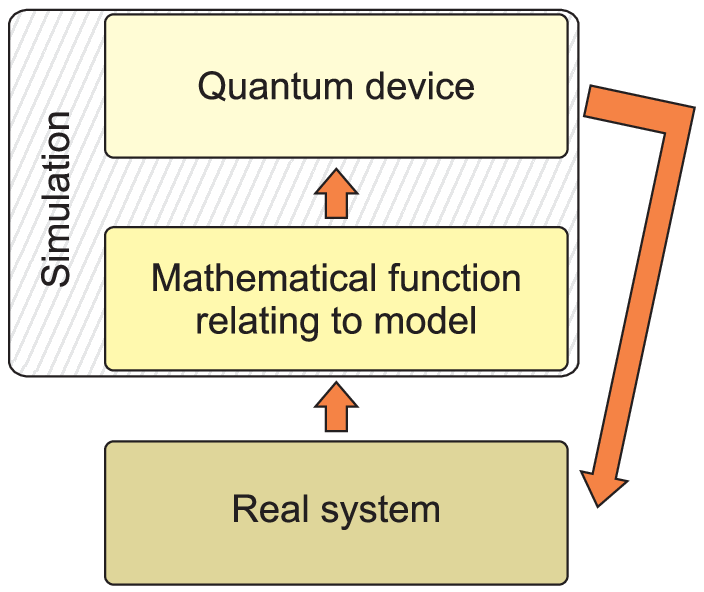}
\end{center}
\caption{A quantum simulator reveals information about an abstract mathematical function relating to a physical model. However, it is important to consider the typical purpose and context of such a simulation. By comparing its results to a real system of interest, a simulation is used to decide whether or not the model accurately represents that system. If the representation is though to be accurate, the quantum simulator can then loosely be considered as a simulator for the system of interest. We represent this in the figure by a feedback loop from the quantum device back to the system of interest.}
\label{fig:fig1}
\end{figure}
Both simulators and computers are physical devices that reveal information about a mathematical function. Whether we call a device a simulator or a computer depends not only on the device, but also on what is supposed about the mathematical function and the intended use of the information obtained. 

If the function is interpreted as part of a physical model then we are likely to call the device a simulator. However, this brief definition neglects the typical purpose and context of a simulation (see \fir{fig:fig1}). As will become clear below, a simulation is usually the first step in a two-step process, with the second being the comparison of the physical model with a real physical system (see \secr{sec:howused} `How are simulators used?'). This then makes simulation part of the usual scientific method. This context is why some loosely state that simulation is the use of one physical device to tell us about another real physical system~\cite{Feynman1982}. It also affects the level of trust that can be reasonably demanded of the simulation (see \secr{sec:whentrust} `When are quantum simulators trustworthy?').

If the accuracy with which a device simulates a model can be arbitrarily controlled and guaranteed then it is often elevated to the status of a computer, a name that reflects our trust in the device. A consequence of this guaranteed accuracy is that it allows assured interpretation of the results of the operation, the information obtained about a mathematical function, without reference to some real system. Thus, as well as to imply accuracy, the term computer is also more often used to describe calculations that relate to more abstract mathematical functions, unconnected to a physical system, and are used outside of the scientific method.

It is interesting to apply our definition of a simulator to well-known situations in which the term is used. The majority of experimental devices advertised as quantum simulations are so-called analogue simulators~\cite{Kendon2010,Cirac2012,Hauke2012,Schaetz2013,Buluta2009,Georgescu2013}. They are devices whose Hamiltonians can be engineered to approximate those of a subset of models put forward to describe a real system. This closely fits our definition of simulators, and their usual purpose and context outlined above. Another different type of device is Lloyd's digital quantum simulator~\cite{Lloyd1996}. This replicates universal unitary evolution by mapping it, via Trotter decompositions, to a circuit, which can then be made arbitrarily accurate by the use of error correction. Whilst going by the name simulator, it is effectively a universal quantum computer. From our arguments above, we would also describe this as a computer: error correction ensures the result applying to the modelled system can be interpreted without comparison to a real physical system, thus playing the role of a computation. Finally, the company D-wave has developed a device to find the ground state of the classical Ising model~\cite{Johnson2011b}. While this is a device that returns a property of a physical model, it is advertised as a computer. We would agree, since its primary use seems to be in solving optimisation problems embedded in the Ising ground state, rather than learning about a real physical system.

\section{What are quantum simulators?}
\label{sec:whatquant}
To complete the definition of a {\em quantum} simulator we need to define what is meant by a quantum device.
This problem is also faced by quantum biology \cite{Davies2004,Abbott2008,Lambert2013} and other quantum technologies. It is complicated by the fact that, at some level, quantum mechanics describes the structure and dynamics of all physical objects. Quantumness may be structural and inert e.g.\ merely responsible for the available single-particle modes. Or quantumness may be active e.g.\ exploiting entanglement between modes, potentially achieving functionality more efficiently than a classical device (see \secr{sec:whyneedquant} `Why do we need quantum simulators?').

To this end, we must distinguish between devices for which, during the operation of the simulator, the particular degrees of freedom doing the simulating do or do not behave classically. We choose here to define classical as when there is some single-particle basis in which the density operator $\hat{\rho}(t)$ describing the relevant degrees of freedom is, for the purposes of the simulation, diagonal at all times $t$. This is written
\begin{equation}
\hat{\rho}(t) = \sum_{s,i} p( \left \{ N_{s,i} \right \} , t) \ket{\left \{ N_{s,i} \right \} ,t} \bra{\left \{ N_{s,i} \right \} ,t} . \nonumber
\end{equation}
Here $\ket{\left \{ N_{s,i} \right \} ,t}$ is a Fock state in which $N_{s,i}$ particles of species $s$ occupy mode $i$. The mode annihilation operator is $\an{a}_{s,i}(t) = \int \dd \rrr \an{\Psi}_s (\rrr) \chi^\ast_{s,i} (\rrr,t)$, with $\chi_{s,i} (\rrr,t)$ the corresponding single-particle modefunction and $\an{\Psi}_s (\rrr)$ the field operator for species $s$. The diagonal elements $p( \left \{ N_{s,i} \right \} , t)$ are the probabilities of the different occupations.

This condition ensures there is always a single-particle basis in which dephasing would have no effect. This invariance under dephasing is a common way to define classicality~\cite{Meznaric2013}. The condition also disallows entanglement between different single-particle modes, as would be expected for a condition of classicality. It does allow the natural entanglement between identical particles in the same mode due to symmetrisation. Such entanglement can be mapped to entanglement between modes by operations that themselves do not contribute entanglement~\cite{Killoran2014}. However, if such operations are never applied, it is reasonable to consider the device to be classical. In other words, we are less concerned with the potential of entanglement as a resource than how this resource is manifested during the operation of the device.

Let us build confidence in our definition by using it to classify well-known devices as classical or quantum. Reassuringly, the operation of the room-temperature semi-conductor devices used to perform every-day computing are classical according to the definition. The relevant properties of inhomogeneous semi-conductors are captured by a model in which the degrees of freedom are valence (quasi) electrons that incoherently occupy single-particle states $\chi_i (\rr)$ of the Bloch type~\cite{Ashcroft1976}. Next, consider two devices for preparing the ground state of a classical Ising model, classical annealing~\cite{Kirkpatrick1983} and quantum annealing~\cite{Farhi1998,Farhi2000,Santoro2006,Das2008,Biamonte2011b}. Classical annealing by coupling the Ising spins to a cooling environment is not quantum since at all times the thermal density matrix of the system is diagonal in the computational basis, a single-particle basis. However, preparing that same state by quantum annealing, adiabatically quenching a transverse field, is expected to be quantum. This is due to the fact that in the middle of the quench, which forms the main part of the simulation, the Ising spins will usually become entangled. Since these are particles in distinguishable modes, the device cannot behave classically at all times. Finally, consider a Bose-Einstein condensate~\cite{Leggett2001,Pitaevskii2003} that is accurately described by many bosons in the same single-particle mode $\chi_0 (t)$. Alternatively, consider a Poissonian mixture of different occupation numbers or equivalently a coherent number superposition of unknown phase, both of which are well approximated by $N_0$ bosons occupying $\chi_0 (t)$, for large mean occupation $N_0$. In these cases, the single occupied modefunction evolves according to the Gross-Pitaevskii equation. Devices based on this condensate evolution have been proposed to simulate some gravitational models~\cite{Garay2000}, and we would accordingly label these simulations as classical. However, simulating a related model featuring cosmological particle production by exploiting coherent Bogoliubov excitations above the condensate is sensitive to off-diagonal elements in the density operator, in a single-particle basis, and is thus quantum \cite{Jain2007}.

Our chosen boundary between quantum and classical is one of many possibilities; and indeed defining the quantumness of the simulation entirely in terms of the device is not common. Many others~\cite{Buluta2009,Georgescu2013} take the quantum in quantum simulator to relate to the model being simulated as well as to the simulating device. In common with definitions of quantum computation, our assignment of the quantum in quantum simulator based only on the device avoids the assumption that only simulating quantum models is hard enough to potentially benefit from a quantum device. This is not so: finding the ground state of even a classical Ising model is NP-hard and thus thought to be inefficient on both a classical and quantum device~\cite{Barahona1982,Bernstein1997}.

\section{How are simulators used?}
\label{sec:howused}
A common perception (that goes right back to the language used at the conception of quantum simulation~\cite{Feynman1982}) is that the purpose of a simulator is purely to reveal information about another real system. We pick an idealised model describing a system of interest, and then simulate that model, taking the output to describe not only the model but the system of interest. As long as the idealised model is a `good' description of the system of interest then it is inferred that the simulator is a `good' simulator of the system. While this inference is correct, it misses an important purpose of a simulator.

This other crucial purpose of a simulator is to reveal information about a model and compare this to the behaviour of the real system of interest. This then allows us to infer whether or not the model provides a `good' description of the system in the first place and whether or not the results bear any relevance to the real world. For example, simulating the Fermi-Hubbard model would be hugely important if it turned out that this model captures the behaviour of some high-$T_c$ superconductors (as suggested by some~\cite{Anderson1987,Rice1988,Anderson2004,LeHur2009}), but it may be that the main conclusion of simulations will be to rule this out (as expected by others~\cite{Laughlin2002,Leggett2006,Laughlin2014}). Only when we have developed confidence in a model accurately representing a system can we use the simulator of the model to inform us about the system. 

\section{Why do we need simulators?}
\label{sec:whyneed}
Above we have stated that simulators are used to find properties of a model, assess whether the model is relevant to and accurately describes the real system of interest, and, if so, learn about that system.
Are there other ways to learn about a system without simulation? Do we need simulators?

There are, of course, many examples of scientists making progress without simulation. Over a century ago, the phenomenon of superconductivity was discovered and later its properties analysed by experimental investigation largely unguided by analytical or numerical simulation~\cite{Ginzburg2004}. Today, in cases where detailed simulation is not possible, we successfully design drugs largely by trial and error on a mass scale~\cite{Madsen2002}. 

These two examples, however, also show why simulation is crucial. Computer-aided drug design~\cite{Cohen1996,Zhou2010} exploits the simulation of molecular systems to drastically speed up and thus lower the cost of the design process. Similarly, if we wish to manufacture materials with enhanced superconducting properties, e.g. increase the critical temperature $T_c$, then we might benefit from some understanding directing that manufacture, as would be provided by a model and a means of simulating it~\cite{Fausti2011,Kaiser2012}. 

Simulation can also be a convenience: in 2014 the USA bobsleigh team won Olympic bronze with a machine designed almost entirely virtually~\cite{Bobsleigh}. Simulation was used to optimise the aerodynamic performance without the need for a wind tunnel.

\section{Why do we need quantum simulators?}
\label{sec:whyneedquant}
While the idea of simulations is centuries old \cite{Brewster1830,Thomson1876}, the suggestion that a quantum device would make for a better mimic of some models than a classical device is commonly attributed to Feynman in 1982 \cite{Feynman1982}. He noted that calculating properties of an arbitrary quantum model on a classical device is a seemingly very inefficient thing to do (taking a time that scales exponentially with the number of particles in the model being simulated), but a quantum device might be able to do this efficiently (taking a time that scales at most polynomially with particle number~\cite{Lloyd1996}).

This does not of course prohibit the simulation of many quantum models from being easy using classical devices and thus not in need of a quantum simulator. The classical numerical tools usually employed include exact calculations, mean-field~\cite{Stanley1971} and dynamical mean-field theory~\cite{Georges1996,Kotliar2006,Aoki2013}, tensor network theory~\cite{Verstraete2008,Cirac2009,Johnson2010,Schollwock2011,Biamonte2011,Evenbly2011,Johnson2013,Orus2013}, density functional theory (DFT)~\cite{Hohenberg1964,Kohn1965,Parr1983,Martin2004,Burke2005} or quantum Monte Carlo algorithms~\cite{Foulkes2001,Gull2011,Pollet2012,Austin2012}, which all have their limitations. Exact calculations are only possible for small Hilbert spaces. Mean-field-based methods are only applicable when the correlations between the constituent parts of the system being modelled are weak. Tensor network methods are only applicable if there is a network structure to the Hilbert space and often fail in the presence of strong entanglement between contiguous bipartite subspaces~\cite{Eisert2010}, with this sensitivity to entanglement being much greater with two- or higher-dimensional models. For DFT, the functionals describing strong correlations are, in general, not believed to be efficient to find~\cite{Schuch2009}. Quantum Monte Carlo struggles, for example, with Fermionic statistics or frustrated models, due to the sign problem~\cite{Loh1990,Troyer2005}. 

For the above reasons, quantum devices are expected to be crucial for large network (e.g. lattice) models, featuring Fermions or frustration and strong entanglement, or non-network based many-body models featuring states with strong correlations that are difficult to describe with DFT. Strong entanglement can arise, for example, near a phase transition, or after a non-equilibrium evolution~\cite{Trotzky2012}. It must be stated, however, that there is no guarantee that a classical device or algorithm will not sometime in the future be devised to efficiently study some subset of the above quantum models. 

In addition to the widely-accepted need for quantum devices for the quantum models discussed above, there are calls and proposals for quantum devices to simulate classical models~\cite{Yung2010,Sinha2010}, for example, molecular dynamics~\cite{Harris2010} and lattice gas models~\cite{Boghosian1998,Meyer2002}. This also applies to any simulation that reduces to solving an eigenvalue equation~\cite{Abrams1999} or a set of linear equations~\cite{Harrow2009}. As with quantum models, many of these simulations, for example solving a set of linear equations, can be solved without much trouble on a classical device for small to medium simulations. The benefit of a quantum device is that the size of problems that can be tackled in a reasonable time grows significantly more quickly with the size of the simulating device than it does for a classical device, thus it is envisaged that quantum devices will one day be able to solve larger problems than their classical counterparts.

It is clear from this last point that the scaling of classical and quantum simulators must be treated carefully, taking into account the sizes of problems that can be tackled by current or future devices. It is possible that the experimental difficulty of scaling up quantum simulation hardware might cause an overhead such that a quantum device does not surpass the accuracy obtained by a classical algorithm that in principle does not scale as well but runs on ever-improving hardware obeying Moore's law.

\section{When are quantum simulators trustworthy?}
\label{sec:whentrust}
\begin{figure*}[tbp]
\begin{center}
  \includegraphics[width=121mm]{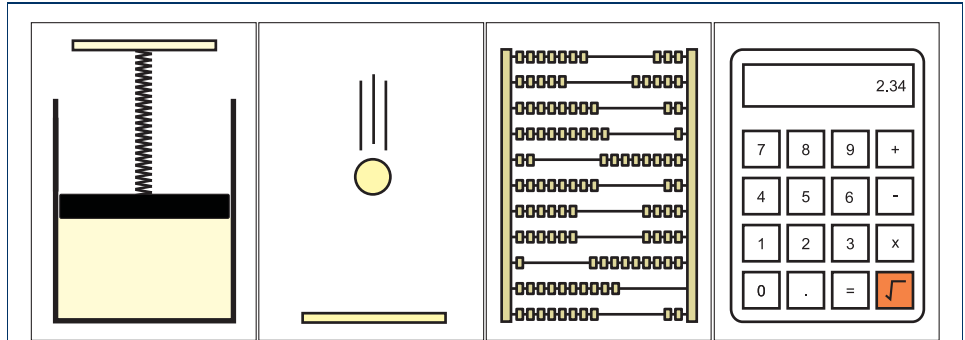}
\end{center}
\caption{Consider the displacement of a spring due to the pressure of a gas (far left), or the time taken for a dropped ball to fall (middle left). Simple models can be proposed to describe either system. The former might be modelled as an ideal gas trapped in a box by a frictionless piston held in place by a perfect spring. The latter as a frictionless body moving with uniform acceleration. Calculating the quantity of interest within either system, displacement or time, respectively, reduces within the model to calculating a square root. We thus consider four methods to perform a this simulation. Building an approximation to either model system; analogue simulation. Alternatively, using an abacus (middle right) or a calculator (far right); digital simulation. \newline \newline
With today's knowledge, in the parlance used in this article, we would elevate the status of the latter two simulations to computations, because of the guaranteed accuracy with which each calculation reproduces the model. Meanwhile, the former two simulatiors are not so easily verified. Importantly, they are falsifiable, e.g.\ by comparing one to the other. This is similar to the state of analogue quantum simulators currently used to perform large-scale quantum simulations.  \newline \newline
However, the confidence in each simulator is a matter of perspective. It is not objective. Many centuries ago, we would only have trusted the abacus to perform such a calculation, since its principles were well understood and square-root algorithms with assured convergence were known even to the Babylonians.
Once Gallileo began the development of mechanics, we might have considered the method of dropping a ball. Confidence in the simulation could have been established by testing the analogue simulator against the abacus. Nearly two centuries ago, when we first began to understand equilibrium thermodynamics, we might have preferred the gas-piston-spring method. Nowadays, we would all choose the calculator or a solid-state equivalent. This confidence is partly a result of testing the calculator against some known results, but also largely because, after the development of quantum mechanics, we feel we understand the components of solid-state systems to such a high level that we are willing to extrapolate this confidence to unknown territory. In a century, our confidence could well be placed most strongly in another system.
}
\label{fig:fig2}
\end{figure*}
So far we are yet to address perhaps the most difficult and important aspect of simulation, upon which its success rests. How can we asses whether the quantum simulator represents the model? How rigorous an assessment is needed?

For this discussion we focus on analogue quantum simulators, because they are the most easily scaled quantum simulators and so are likely to be used in the near future to simulate large systems. They also most closely follow our definition of a simulator, as opposed to a computer (see \secr{sec:whatsim} `What are simulators?').

The topic of falsifying bad quantum simulators has received some attention. In certain parameter regimes there may be efficiently calculable exact analytical results or it might be possible to perform a trusted classical simulation, against which the quantum simulator results may be compared~\cite{Trotzky2012}. Often there are bounds that some measurable quantities are known to obey, and this too can be tested~\cite{Hauke2012}. Alternatively, it might be possible to check known relationships between two different simulations. For example, in an Ising model, flipping the direction of the magnetic field is equivalent to flipping the sign of the component of the spins along that field, thus giving two simulations whose results are expected to have a clear relationship. A natural extension of this strategy is to compare many quantum simulations realised by different devices, perhaps each with a slightly different source of error, trusting only the aspects of the results shared by all devices~\cite{Leibfried2010}. If any of the above tests fail beyond an acceptable accuracy, then we do not trust the simulation results. If a simulator passes all tests, then we may take this as support for the accuracy of that simulator.

It would be incorrect, however, to say that such tests verify the accuracy of a simulator. A simulator could have significant errors yet pass these tests. It might be that the simulator is accurate in the regimes in which we have accurate analytical or classical numerical results, but is more sensitive to errors in regimes that are difficult to treat with other methods, e.g.\ near phase transitions, perhaps for the same reason. In fact, Hauke {\em et al.}\ gave an example of exactly this phenomenon in the transverse Ising model~\cite{Hauke2012}. The danger with comparing simulations, even realised by different devices~\cite{Leibfried2010}, is that there may be similar sources of error, or errors in the two simulations may manifest in the results in the same way.
 
Although this makes simulation difficult to assess, it does not invalidate it; it would be unreasonably harsh to demand verification of all simulators. The reason for this is that, as illustrated in \fir{fig:fig1}, simulators are usually the first step in a two-step process: first a device is devised to simulate a model, and second the model is employed to study a real system (see \secr{sec:howused} `How are they used?'). It might be unreasonable to demand a more rigorous testing of the first part of this process than the second.
In the second part, when we devise a model to reproduce the behaviour of a physical system, we only demand that the model be falsifiable~\cite{Popper1963}. We seek as many fail-able tests as possible of the model, and to the extent that it passes these tests, we retain the model. It is difficult for experiments to verify a particular use of the model, rather successful experiments merely declare the model `not yet false'. This is the scientific method. We should not, therefore, demand anything more or less when going in the other direction, devising a physical device to reproduce the behaviour of a model. All we can do is test our simulators as much as possible, and slowly build confidence in accordance with the passing of these tests. If the capability of performing such tests lags behind the development of the simulator, then so naturally must our confidence.

It becomes clear that the purpose of the device is crucial to how it is assessed, explaining our highlighting the purpose of a simulator alongside its definition. If we were using a device to provide information about a model without any additional motivation, as with a computer, then it would be reasonable to search for a means of verification and guarantees of accuracy, as with a computer. Eventually, quantum technologies might develop to a stage where large simulations of this type are feasible, e.g.\ via Lloyd's digital simulator~\cite{Lloyd1996}, but it is likely to be in the more distant future. It must be noted, however, that many of the devices we use regularly for computation are unverifiable in the strictest sense. Not every transistor in the classical computers we use (for instance to simulate quantum systems) can be verified to be functioning as desired~\cite{May1992}. We instead develop an understanding of the sources of error, perform some tests to check for obvious errors, and use the devices with caution.

The words `trust' and `confidence' in the preceding paragraphs are chosen deliberately. They indicate that, since for simulation we do not always have verifiability, we are not discussing objective properties of devices, but our understanding of them. This will change in time (see an example of this in \fir{fig:fig2}). Further, confidence depends on the eventual goal of our use of the simulator. Some properties of a system may be too sensitive to Hamiltonian parameters to be realistically captured by a simulator, while other properties may be statistically robust against parameter variations~\cite{Walschaers2013}. In this sense trustworthiness is not a clear-cut topic that is established upon the initial development of a simulator. Instead, it is the result of a complex, time-consuming process in the period that follows. It is the responsibility of critics not to be overly harsh and unfairly demanding of new simulators to provide immediate proof of their trustworthiness, but it is also the responsibility of proponents not to declare trustworthiness before their simulator has earned it.

\section{Where next for quantum simulation?}
\label{sec:wherenext}
The majority of the current effort on quantum simulation is, firstly, in matching models of interest to a suitable quantum device with which to perform a simulation~\cite{Buluta2009,Georgescu2013}. Secondly, experimentalists demonstrate a high level of control and flexibility with a simulator, performing some of the simple fail-able tests mentioned above~\cite{Aspuru-Guzik2012,Blatt2012,Bloch2012}. This is very much along the lines of the five
goals set out by Cirac and Zoller in 2012 \cite{Cirac2012}, and great successes have led to claims that we are now able to perform simulations on a quantum device that we are unable to do on a classical device. In the future, the main direction of inquiry will continue to be along these lines.

However, it is the very fact that the simulation capabilities of quantum devices are beginning to surpass those of classical devices that should prompt a more forceful investigation into the best approach to establishing confidence in quantum simulators. Hauke {\em et al.}\ proposed a set of requirements for a quantum simulator, an alternative to Cirac and Zoller's, that focuses on establishing the reliability and efficiency of a simulator, and the connection between these two properties \cite{Hauke2012}. As we move to classically unsimulable system sizes and regimes where there is no clear expected behaviour, trustworthiness and falsifiability should no longer be an afterthought. In fact, they should be primary objectives of experimental and theoretical work, since quantum simulators cannot truly be useful until some level of trust is established.

 Can we predict in advance where the results of quantum simulators are more sensitive to errors? How does this overlap with the regimes of classical simulability? Are there even some results that will be exponentially sensitive to the Hamiltonian parameters and not expected to ever be simulable in a strict sense? These are difficult but important questions to answer, and progress on them will be exciting and thought provoking.


\begin{backmatter}

\section*{Competing interests}
  The authors declare that they have no competing interests.

\section*{Author's contributions}
All authors conceived of the study, participated in its design and coordination, helped to draft the manuscript, and read and approved the final manuscript. THJ undertook the majority of the writing.

\section*{Acknowledgements}
The authors thank the National Research Foundation
and the Ministry of Education of Singapore for support.

\bibliographystyle{bmc-mathphys} 
\bibliography{paper}      

\end{backmatter}
\end{document}